\documentclass[twocolumn,showpacs,preprintnumbers,amsmath,amssymb,floatfix,nofootinbib]{revtex4}
\usepackage{amsmath,amsfonts,bbm,euscript,hyperref}
\usepackage{graphicx}
\usepackage{dcolumn}
\usepackage{bm}
\usepackage{epsfig}
\epsfclipon

\newcommand{\nco}{\newcommand}
\nco{\beq}{\begin{equation}} \nco{\eeq}{\end{equation}}
\nco{\beqa}{\begin{eqnarray}} \nco{\eeqa}{\end{eqnarray}}
\def\be{\begin{equation}}
\def\ee{\end{equation}}    
\def\baray{\begin{eqnarray}}
\def\earay{\end{eqnarray}}

\nco{\lra}{\leftrightarrow}

\nco{\sss}{\scriptscriptstyle} \nco{\dphi}{\varphi}
\nco{\lsim}{\mbox{\raisebox{-.6ex}{~$\stackrel{<}{\sim}$~}}}
\nco{\gsim}{\mbox{\raisebox{-.6ex}{~$\stackrel{>}{\sim}$~}}}

\def\IK{\relax{\rm I\kern-.20em K}}
\def\IM{\relax{\rm I\kern-.20em M}}

\def\lsim{\mbox{\raisebox{-.6ex}{~$\stackrel{<}{\sim}$~}}}
\def\gsim{\mbox{\raisebox{-.6ex}{~$\stackrel{>}{\sim}$~}}}
\def\sss{\scriptscriptstyle}

\begin{document}

\preprint{UMN-TH-2926/10}

\title{Large Nongaussianity in Axion Inflation}

\author{Neil Barnaby, Marco Peloso}

\affiliation{%
\centerline{School of Physics and Astronomy, University of Minnesota, Minneapolis, MN 55455, USA}
e-mail:\ barnaby@physics.umn.edu, peloso@physics.umn.edu}


\begin{abstract} 
The inflationary paradigm has enjoyed phenomenological success, however, a compelling particle physics realization is still lacking.  The key obstruction is that the requirement of a suitably flat scalar potential is 
sensitive to Ultra-Violet (UV) physics.  Axions are among the best-motivated inflaton candidates, since the flatness of their potential is naturally protected by a shift symmetry.  We re-consider the cosmological 
perturbations in axion inflation, consistently accounting for the coupling to gauge fields $\phi F \tilde{F}$, which is generically present in these models. This coupling leads to production of gauge quanta, 
which provide a new source of inflaton fluctuations, $\delta\phi$. For an axion decay constant  $\lsim 10^{-2} \,  M_p$, this effect typically dominates over the standard fluctuations from the vacuum, 
and saturates the current observational bounds on nongaussianity of the CMB anisotropies. Since sub-Planckian values of the decay constant are typical for concrete realizations that admit a UV completion 
(such as N-flation and axion monodromy), we conclude that  large nongaussianity is easily obtained in very minimal and natural realizations of inflation.
\end{abstract}


\maketitle

\section{Introduction} 
\vspace{-4mm}

Primordial inflation is the dominant paradigm  in current cosmology since (i) it resolves the conceptual difficulties of the standard big bang model, and (ii) it predicts primordial perturbations with properties in 
excellent agreement with those that characterize the Cosmic Microwave Background (CMB) anisotropies.  
Despite these successes, there is still no compelling particle physics model of inflation; the key obstacle being the requirement of a sufficiently flat scalar potential, $V(\phi)$.  Even generic Planck-suppressed 
corrections 
may yield unacceptably large contributions to the slow roll parameters $\epsilon\equiv\frac{M_p^2}{2}\left(\frac{V'}{V}\right)^2$, $\eta\equiv M_p^2 \frac{V''}{V}$, thus spoiling inflation (prime denotes derivative 
with respect to $\phi$, while $M_p\cong 2.4\cdot 10^{18}\mathrm{GeV}$ is the reduced Planck mass). One of the simplest solutions to this problem is to assume that the inflaton $\phi$ is a 
Pseudo Nambu Goldstone Boson (PNGB) \cite{natural}-\cite{kaloper}.  In this case the inflaton enjoys a shift symmetry $\phi\rightarrow \phi + \mathrm{const}$, which is broken either explicitly 
or by quantum effects.  In the limit of exact symmetry, the $\phi$ direction is flat and thus dangerous corrections to $\epsilon$, $\eta$ are controlled by the smallness of the symmetry breaking.  
Moreover, PNGBs like the axion are ubiquitous in particle physics: they arise whenever an approximate global symmetry is spontaneously broken and are plentiful in string compactifications.  Axion 
inflation is also phenomenologically desirable since the tensor-to-scalar ratio is typically large in such models.

The first explicit example of axion inflation was the natural inflation model \cite{natural} in which the shift symmetry is broken down to a discrete subgroup $\phi \rightarrow \phi + (2\pi)f$,
resulting in a periodic potential 
\begin{equation}
\label{Vnp}
  V_{\mathrm{np}}(\phi) \cong \Lambda^4 \left[1-\cos\left(\phi / f\right)\right]
\end{equation}
with $f$ the axion decay constant.   For such potential, agreement with observations requires $f > M_p$, which may be problematic since it suggests a global symmetry broken above the quantum gravity scale, where 
effective field theory
is presumably not valid.  Moreover, $f > M_p$ does not seem possible in string theory \cite{big_f}.  More recently, several controlled realizations of axion inflation have been studied --  including 
double-axion inflation \cite{2-flation}, N-flation \cite{N-flation,N-flation2}, axion monodromy \cite{monodromy} and axion/4-form mixing \cite{kaloper} -- which have $f< M_p$ but nevertheless behave effectively 
as large field inflation models 
($\phi \gsim M_p$).

In axion inflation models, the inflaton couples to some gauge field as $\frac{\alpha}{f} \phi F^{\mu\nu} \tilde{F}_{\mu\nu}$, where $F_{\mu \nu}=\partial_\mu A_\nu - \partial_\nu A_\mu$  
and $\tilde{F}^{\mu\nu} = \epsilon^{\mu\nu\alpha\beta} F_{\alpha\beta}/2$.  The scale of this coupling is set by  the axion decay constant, $f$; the dimensionless parameter $\alpha$ is typically
order unity but can be $\geq 1$ in multi-field \cite{2-flation}  or extra-dimensional models \cite{lorenzo}.  It is natural to explore the implications of this generic interaction for observables.  In 
\cite{lorenzo} it was shown that energy dissipation into gauge fields can slow the motion of $\phi$, providing a novel new inflationary mechanism that operates at very strong coupling.  Here, we point out that 
even in the conventional slow-roll regime, the coupling $\phi F\tilde{F}$ can have significant impact.  The motion of the inflaton amplifies the fluctuations 
of the gauge field, which in turn produce inflaton fluctuations via \emph{inverse decay} \cite{ID}: $\delta A+\delta A\rightarrow \delta\phi$.  When $f \lsim 10^{-2} \, M_p$, which is  natural for 
realizations that admit an UV completion, we show that the inverse decay typically dominates over the usual vacuum fluctuations from inflation, and this has dramatic phenomenological consequences.  Our results
are quite general: in the spirit of effective field theory, a coupling $\phi F \tilde{F}$ should be included whenever $\phi$ is pseudo-scalar \cite{EFT}.

Recently, there has been considerable interest in nongaussian effects in the CMB (see the review \cite{NGreview} for references).  Nongaussianity will be probed to unprecedented accuracy 
with the forthcoming Planck data and may provide a valuable tool to discriminate between models.  Several constructions are known which can predict an observable signature;  
however, in the minimal cases nongaussianity is small, and obtaining an observable level usually requires either fine-tuning or unconventional field theories.  Here we point out that the inverse decay 
contribution to $\delta\phi$ is highly nongaussian in axion models; observational bounds are easily saturated for $f$ not much smaller than $M_p$.  Thus, the simplest and, perhaps, most natural models 
of inflation can  lead to observable nongaussianity.  

\vspace{-4mm}
\section{Cosmological Perturbations}
\vspace{-4mm}

We consider the theory
\begin{equation}
\label{L}
  S = -\frac{1}{2}(\partial\phi)^2 - V(\phi) - \frac{1}{4}F^{\mu\nu}F_{\mu\nu} - \frac{\alpha}{4 f} \phi F^{\mu\nu}\tilde{F}_{\mu\nu}
\end{equation}
where $\phi$ is the PNGB inflaton, $F_{\mu\nu}$ the field strength of the gauge field (for simplicity, a $U(1)$ gauge field is considered; the extension to non-Abelian groups is
straightforward), and $\tilde{F}_{\mu\nu}$ its dual.  The potential $V(\phi)$ may contain a periodic contribution of the form (\ref{Vnp}) due to non-perturbative effects and, perhaps, non-periodic 
contributions from other effects (such as moduli stabilization or wrapped branes). In this Section, we leave $V(\phi)$ arbitrary, except to suppose that it is sufficiently flat to support 
$N_e \gsim 60$ $e$-foldings of inflation.  We assume an FRW geometry $ds^2=-dt^2 + a(t)^2 d{\bf x}^2 = a(\tau)^2\left[-d\tau^2 + d{\bf x}^2\right]$.

Working in Coulomb gauge, we decompose $\vec{A}(t,{\bf x})$ into circular polarization modes obeying \cite{lorenzo}
\begin{equation}
\label{gauge_eom}
  \left[\frac{\partial^2}{\partial\tau^2} + k^2 \pm \frac{2 k \xi}{\tau} \right] A_{\pm}(\tau,k) = 0, \hspace{3mm}\xi \equiv \frac{\alpha \dot{\phi}}{2 f H}
\end{equation}
where dot denotes differentiation with respect to $t$, $H \equiv \dot{a} / a$, $\xi \cong \mathrm{const}$.  We observe that one of the polarizations of $\vec{A}$ 
experiences a tachyonic instability for $k/(a H) \lsim 2\xi$.  
The growth of fluctuations is described by \cite{lorenzo}
\begin{equation}
\label{tac}
  A_{+}(\tau,k)\cong \frac{1}{\sqrt{2k}}\left(\frac{k}{2\xi a H}\right)^{1/4} e^{\pi \xi - 2\sqrt{2\xi k / (a H)}}
\end{equation}
in the interval $(8\xi)^{-1} \lsim k/(aH) \lsim 2\xi$ of phase space which accounts for most of the power in 
the produced gauge field (we take $\dot{\phi} > 0$ without loss of generality).  This interval is nonvanishing only for  $\xi \gsim \mathcal{O}(1)$, which we assume in the following. 
The production is uninteresting at smaller $\xi$.

The unstable growth of $A_{+}(\tau,k)$ yields an important new source of cosmological fluctuations, $\delta\phi$.  The perturbations of the inflaton are described by \cite{lorenzo,rescattering}
\begin{equation}
\label{inf_eom}
  \left[ \frac{\partial^2}{\partial t^2}+ 3 H \frac{\partial}{\partial t} - \frac{\nabla^2}{a^2} \right] \delta \phi(t,{\bf x}) = \frac{\alpha}{f} F^{\mu\nu}\tilde{F}_{\mu\nu}
\end{equation}
where the source term is constructed from (\ref{tac}).  The solution of (\ref{inf_eom}) splits into two parts: the solution of the homogeneous equation and the particular solution which is 
due to the source.  Schematically, we have
\begin{equation}
\label{hom+par}
  \delta \phi = \underbrace{\delta\phi_{\mathrm{vac}}}_{\mathrm{homogeneous}} + \underbrace{\delta\phi_{\mathrm{inv.decay}}}_{\mathrm{particular}}
\end{equation}

The quantity of interest is the primordial curvature perturbation on uniform density hypersurfaces, $\zeta = -\frac{H}{\dot{\phi}} \delta\phi$.  We computed the two-point 
$\langle \zeta \left( {\bf x} \right) \zeta \left( {\bf y} \right) \rangle$ and three-point $\langle \zeta \left( {\bf x} \right) \zeta \left( {\bf y} \right) \zeta \left( {\bf z} \right) \rangle$
correlation functions using the formalism of \cite{lorenzo,rescattering}.  The two-point function defines the power spectrum
\begin{equation}
  \langle \zeta({\bf x}) \zeta({\bf y}) \rangle = \int \frac{dk}{k}\, \frac{\sin\left[k|{\bf x}-{\bf y}|\right]}{k|{\bf x} - {\bf y}|} \, P_\zeta(k)
\end{equation}
We find the result 
\begin{equation}
\label{pwr}
  P_{\zeta}(k) =
     \mathcal{P} \left(\frac{k}{k_0}\right)^{n_s-1}\left[ 1 + 7.5\cdot 10^{-5} \, \mathcal{P} \, \frac{e^{4\pi\xi}}{\xi^6} \right] 
\end{equation}
where $\mathcal{P}^{1/2} \equiv   \frac{H^2}{2\pi |\dot{\phi}|} $,  $n_s$ is the spectral index, and  the pivot scale is $k_0 = 0.002 \, {\rm Mpc}^{-1}$. The two terms in (\ref{pwr}) are the power spectra of 
the homogeneous and inhomogeneous parts of (\ref{hom+par}), respectively. 
There is no ``mixed term'' since the two contributions (\ref{hom+par}) are uncorrelated. (The gauge fluctuations that source $\delta\phi_{\mathrm{inv.decay}}$, and that are amplified according to (\ref{gauge_eom}),  
are not correlated with the vacuum inflaton fluctuations.)  The power spectrum is probed by CMB and Large Scale Structure observations. It is found to be nearly scale invariant ($n_s \simeq 1$, the precise value 
depends on the data set assumed \cite{WMAP7}), and have amplitude $P_\zeta(k) \cong 25\cdot 10^{-10}$ 
\cite{CMBPol} (the so-called COBE normalization).  When inverse decay fluctuations are subdominant, we have the 
standard result $\mathcal{P}^{1/2} = 5\cdot 10^{-5}$; however, at large $\xi$ the value of $\mathcal{P}$ must be modified.

The three-point correlation function encodes departures from gaussianity.  Nongaussian effects from inverse decays are maximal when all three modes have comparable wavelength (the equilateral configuration).  
The magnitude of the three-point function is conventionally quantified using the parameter $f_{NL}$ \cite{WMAP7}.  We find that:
\begin{equation}
\label{fNL}
  f_{NL}^{\mathrm{equil}} \cong 4.4\cdot 10^{10}\, \mathcal{P}^3\, \frac{e^{6\pi\xi}}{\xi^9}
\end{equation}
This result does not include the negligible contribution from $\delta\phi_{\mathrm{vac}}$ and is accurate as long as $|f_{NL}|\gsim 1$. From eqs. (\ref{pwr}) and (\ref{fNL}) we can see that 
$f_{NL}^{\mathrm{equil}}  \simeq 8400$ at large $\xi$. The current WMAP bounds are $-214 < f_{NL}^{\mathrm{equil}} < 266$ ($95\%$ CL), while the Planck satellite, and planned missions, will constrain 
$f_{NL}^{\mathrm{equil}}$ to 
$\mathcal{O}(10)$ \cite{Yadav:2010fz}. 

\begin{figure}[h]
\includegraphics[width=0.3\textwidth,angle=-90]{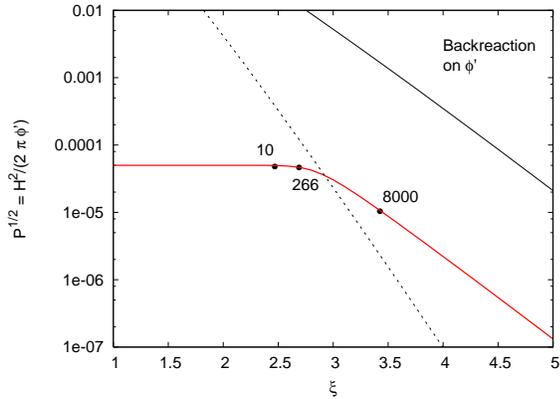}
\caption{Values of parameters  leading to the observed COBE normalization of the power spectrum (red line), and  reference  values for the nongaussianity parameter $f_{NL}^{\mathrm{equil}}=10,266,8000$ 
along this curve. See the main text for details.}
\label{Fig:contours}
\end{figure}

The results (\ref{pwr}) and (\ref{fNL}) only depend on the two dimensionless combinations $\xi$ and $\mathcal{P}^{1/2}$, shown in Figure~\ref{Fig:contours}. The solid red curve indicates the parameter values 
which reproduce the COBE normalization of the power spectrum. In the region below, and above the dashed black line the power spectrum is dominated by $\delta\phi_{\mathrm{vac}}$ and by $\delta\phi_{\mathrm{inv.decay}}$, 
respectively. This second regime is already ruled out by the current bound on $f_{NL}^{\mathrm{equil}}$.

The results (\ref{pwr}) and (\ref{fNL}) have been obtained by disregarding two backreaction effects of the produced gauge quanta. Such quanta are produced at the expense of the kinetic energy of $\phi$, so that, 
if the instability is sufficiently strong, then it will affect the inflaton dynamics. The region of parameter space where this occurs is above the black solid line 
($\mathcal{P}^{1/2} > 13 \xi^{3/2} \, {\rm e}^{-\pi \xi}$) shown in Figure~\ref{Fig:contours}. We have also disregarded the impact of the energy density of the produced quanta on the expansion rate, $H$. This is 
justified provided ${\rm e}^{2 \pi \xi} / \xi^3 \ll 2 \cdot 10^4 M_p^2/H^2$. This constraint is not expressed in terms of  $\xi$ and $\mathcal{P}^{1/2}$, so we have not included it in Figure \ref{Fig:contours}.  
However, it can be studied on a case-by-case basis.

The gauge quanta also source gravity waves (GW). It is customary to normalize the power of GW to that of the density perturbations. Proceeding analogously to the computation of the density perturbations, we find
\begin{equation}
\label{GW}
r \equiv \frac{P_{\rm GW}}{P_\zeta} = 8.1 \cdot 10^{7} \, \frac{H^2}{M_p^2} \left[ 1 + 4.3 \cdot 10^{-7} \, \frac{H^2}{M_p^2} \, \frac{{\rm e}^{4 \pi \xi}}{\xi^6} \right]
\end{equation}
The tensor-to-scalar ratio, $r$, is an important quantity to discriminate between different inflationary models. The current observational limit is $r \lsim 0.2$ \cite{WMAP7}, and activity is underway to probe 
$r \gsim 0.01$ \cite{CMBPol}. 

\vspace{-4mm}
\section{Predictions for Specific Models}
\vspace{-4mm}

\begin{figure}[h]
\includegraphics[width=0.3\textwidth,angle=-90]{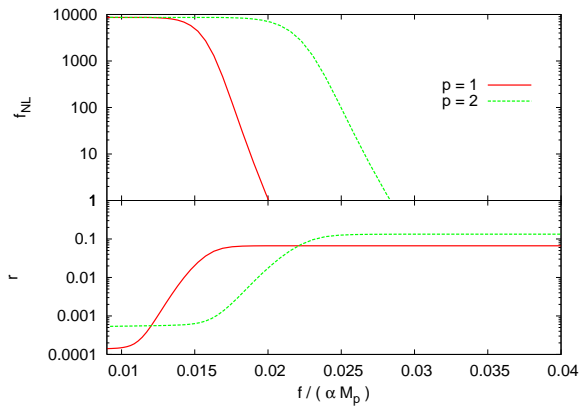}
\caption{Observational predictions for the large-field power-law inflation model (\ref{power_law}) with $p=1,2$ and assuming $N_e\cong 60$.  
The spectral index is $n_s=0.975,0.967$ for $p=1,2$.   At small $f/\alpha$
the coupling of $\phi$ to $F\tilde{F}$ is stronger and nongaussianity is large.  The tensor-to-scalar ratio decreases at strong coupling; however, the decrease is important only at values of $f/\alpha$ 
which are ruled out by the current bound on $f_{NL}^{\mathrm{equil}}$.}
\label{Fig:predictions}
\end{figure}

We now focus our attention on the power-law potential
\begin{equation}
\label{power_law}
  V(\phi) = \mu^{4-p} \phi^p
\end{equation}
which subsumes many interesting scenarios.    Inflation 
proceeds at large field values $\phi \gsim M_p$ and ends when $\phi \sim M_p$.  
For this model, the values of $H$, $\dot{\phi}$ and $n_s$ are uniquely determined by the number of e-foldings of observable inflation $N_e$, according to the standard slow roll inflaton evolution 
($\epsilon ,\, \eta \ll 1$). In the following, we fix $N_e = 60$, which is the typical value taken in large field models. 
Once we do so, we are left with the two parameters $f/\alpha$, and $\mu$. For any given value of 
$f/\alpha$, the mass scale $\mu$ is uniquely determined by fixing the power spectrum (\ref{pwr}) to the COBE value. We can then plot the other observational predictions as a function of $f/\alpha$ only.
We do so in Figure~\ref{Fig:predictions}, where we take $p=1,2$ for illustration. In both cases, backreaction effects can be neglected.

Figure \ref{Fig:predictions} shows that large nongaussianity is rather generic for large-field axion inflation. The current bound is saturated for decay constants $f/\alpha \lsim 10^{-2} M_p$,
which is natural in a model that admits a UV completion.  Current limits on nongaussianity therefore provide an upper bound  on the \emph{strongest} couplings of the type $\phi F\tilde{F}$ between 
the inflaton and \emph{any} gauge field.  

We see also that $r$ decreases at strong coupling.  This modifies the usual predictions of large field inflation and implies, for example, that $p=4$
could be made compatible with observation, at the level of the $2$-point function.  

{\bf Natural Inflation:} The original natural inflation model \cite{natural} was based on the potential (\ref{Vnp}).  If we require $n_s \gsim 0.95$, as suggested by recent data \cite{WMAP7}, then the model 
requires a large decay constant $f \gsim 5 M_p$ \cite{freese3}.  In this regime inflation proceeds near the minimum $\phi=0$ and is indistinguishable from the model (\ref{power_law}) with $p=2$.  
Large values of $f$ weaken the coupling of $\phi$ to $F\tilde{F}$, hence inverse decay is negligible unless $\alpha \gsim 200$, whereas we expect $\alpha = \mathcal{O}(1)$ in the simplest 
(single-axion) scenario.  On the other hand, $f\gsim M_p$ may be problematic and it seems that a UV completion of axion inflation requires $f < M_p$.  We now turn our attention to such 
scenarios.

{\bf Axion Monodromy:}  In \cite{monodromy} an explicit, controlled realization of axion inflation was obtained from string theory.  The potential has the form 
$V(\phi) = \mu^3 \phi + \Lambda^4\cos(\phi/f)$ where the linear contribution arises because the shift symmetry is broken by wrapping an NS$5$-brane on an appropriate 2-cycle, and
the periodic modulation is due to nonperturbative effects.  The former typically dominates \cite{monodromy,monodromy2} so we have the model (\ref{power_law}) with $p=1$, to first 
approximation.  The decay constant is bounded \cite{monodromy} as $0.06 \mathcal{V}^{-1/2} g_s^{1/4} < f/M_p < 0.9 g_s$ with $g_s < 1$ the string coupling and $\mathcal{V} \gg 1$ the 
compactification volume in string units.  From Fig.~\ref{Fig:predictions} we see that large nongaussianity is easily obtained for $\alpha=\mathcal{O}(1)$.  
Periodic modulation of $V(\phi)$ can also lead to resonant nongaussianity \cite{res} for $f\lsim 10^{-2} M_p$ and $\Lambda$ sufficiently large \cite{monodromy2,Hannestad:2009yx}.

{\bf N-flation:} In \cite{N-flation} it was noted that the collective motion of $N$ axions $\phi_i$, each with its own broken shift symmetry,  can support inflation when $f_i < M_p$,
via the assisted inflation  mechanism \cite{assisted}.  This scenario is quite natural in string theory, where generic compactifications may contain exponentially large numbers of axions
\cite{N-flation,N-flation2}.  For $\phi_i \lsim f_i$ we can expand the potential near the minimum to obtain $V \cong \sum_i m_i^2 \phi_i^2 / 2$.  The dynamics of the collective field 
$\Phi \equiv \sqrt{\sum_i\phi_i^2}$ are well-described by a single field model of the form (\ref{power_law}) with $p=2$ \cite{N-flation,N-flation2}.
Sufficient inflation requires $\Phi > M_p$ which can be achieved for sub-Planckian $\phi_i$ provided $\sqrt{N}$ is sufficiently large.

Typically, the mass basis $\{\phi_i\}$ is not aligned with the interaction basis \cite{green,preheat}, so that all $\phi_i$ couple to a given gauge field 
as $\mathcal{L}_{\mathrm{int}} = -\sum_i \alpha_i \phi_i F\tilde{F} / f_i$.  The coupling of the collective field $\Phi$ to $F\tilde{F}$ is highly model-dependent but 
can be parametrized as $\mathcal{L}_{\mathrm{int}} = - \alpha_{\mathrm{eff}} \Phi F\tilde{F} / f_{\mathrm{eff}}$, so that the result of Fig.~\ref{Fig:predictions} apply for the effective
coupling.  A precise calculation depends on the mass rotation and the microphysics, but we expect that an observable signal is possible for reasonable parameters.

{\bf Double-Axion Inflation:} Ref. \cite{2-flation} proposed a model characterized by two axions $\theta$ and $\rho$, with potential
\begin{equation}
V = \sum_{i=1}^2 \Lambda_i^4 \left[ 1 - {\rm cos } \left( \frac{\theta}{f_i} + \frac{\rho}{g_i} \right) \right]
\label{knp}
\end{equation}
which arises from the   coupling of the two axions to two different gauge groups, $\frac{\theta}{f_i} F_i {\tilde F}_i $, and $\frac{\rho}{g_i} F_i {\tilde F}_i $ (up to numerical coefficients).  For 
$f_1/g_1 = f_2/g_2$, one linear combination of the two axions becomes a flat direction of (\ref{knp}). This relation can be ascribed to a symmetry of the theory, and the  curvature of the potential along 
this direction can be made controllably small if this symmetry is only slightly broken. In this case one obtains an effective large field inflaton, with a potential of the type (\ref{Vnp}), and with an 
effective axion constant $> M_p$, even if all the $f_i ,\, g_i$ are sub-Planckian. Therefore, this model can lead to large production of gauge fields and observable $f_{NL}^{\mathrm{equil}}$. 

{\bf Axion Mixing:} Ref.~\cite{kaloper} realizes $p=2$ via axion/4-form mixing.  Here $f<M_p$ so $f_{NL}^{\mathrm{equil}}\gg 1$ is possible.

\smallskip

In summary, we have shown that large nongaussianity is possible for many explicit axion inflation models which admit a UV completion.  Our qualitative results will carry over to any inflation model with
a PNGB dynamically important during inflation, including multi-field models such as the roulette \cite{roulette} or racetrack \cite{racetrack} scenarios.  Similar effects may also be possible for higher
$p$-form fields.  It would be interesting to study the value of $\alpha$ in concrete string theory realizations.

{\bf Acknowledgments } 
We thank L.~Sorbo for insightful comments on an earlier version of this work, and L.~McAllister and R.~Namba for valuable discussions. This work was 
supported in part by DOE grant DE-FG02-94ER-40823 at UMN.


\begin{thebibliography}{99}

\bibitem{natural}

 K.~Freese, J.~A.~Frieman and A.~V.~Olinto,
  Phys.\ Rev.\ Lett.\  {\bf 65}, 3233 (1990).

\bibitem{2-flation}

  J.~E.~Kim, H.~P.~Nilles and M.~Peloso,
  JCAP {\bf 0501}, 005 (2005)
  [hep-ph/0409138].

\bibitem{N-flation}

  S.~Dimopoulos, S.~Kachru, J.~McGreevy and J.~G.~Wacker,
  JCAP {\bf 0808}, 003 (2008)
  [hep-th/0507205].

\bibitem{N-flation2}

  R.~Easther and L.~McAllister,
  JCAP {\bf 0605}, 018 (2006)
  [hep-th/0512102].

\bibitem{monodromy}

  L.~McAllister, E.~Silverstein and A.~Westphal,
  Phys.\ Rev.\  D {\bf 82}, 046003 (2010)
  [arXiv:0808.0706].

\bibitem{monodromy2}

  R.~Flauger, L.~McAllister, E.~Pajer, A.~Westphal and G.~Xu,
  JCAP {\bf 1006}, 009 (2010)
  [arXiv:0907.2916].

\bibitem{lorenzo}

  M.~M.~Anber and L.~Sorbo,
  Phys.\ Rev.\  D {\bf 81}, 043534 (2010)
  [arXiv:0908.4089].

\bibitem{kaloper}
  N.~Kaloper and L.~Sorbo,
  Phys.\ Rev.\ Lett.\  {\bf 102}, 121301 (2009)
  [arXiv:0811.1989].

\bibitem{big_f}

  T.~Banks, M.~Dine, P.~J.~Fox and E.~Gorbatov,
  JCAP {\bf 0306}, 001 (2003)
  [hep-th/0303252].

\bibitem{ID}

The direct decay is instead negligible during inflation: $\Gamma_{\delta\phi\rightarrow\delta A +\delta A} \ll H$.

\bibitem{EFT}

In this case one interprets $M \equiv f/ \alpha$ as the UV scale associated with the validity of the effective description.

\bibitem{NGreview}

  N.~Barnaby,
  Adv.\ Astron.\  {\bf 2010}, 156180 (2010)
  [arXiv:1010.5507].

\bibitem{rescattering}

  N.~Barnaby, Z.~Huang, L.~Kofman and D.~Pogosyan,
  Phys.\ Rev.\  D {\bf 80}, 043501 (2009)
  [arXiv:0902.0615].
  N.~Barnaby,
  arXiv:1006.4615.

\bibitem{WMAP7}

 E.~Komatsu {\it et al.},
  arXiv:1001.4538.

\bibitem{CMBPol}
  D.~Baumann {\it et al.}  [CMBPol Study Team Collaboration],
  AIP Conf.\ Proc.\  {\bf 1141}, 10 (2009)
  [arXiv:0811.3919].

\bibitem{Yadav:2010fz}
  A.~P.~S.~Yadav and B.~D.~Wandelt,
  arXiv:1006.0275.

\bibitem{freese3}

  C.~Savage, K.~Freese and W.~H.~Kinney,
  Phys.\ Rev.\  D {\bf 74}, 123511 (2006)
  [hep-ph/0609144].

\bibitem{assisted}

  A.~R.~Liddle, A.~Mazumdar and F.~E.~Schunck,
  Phys.\ Rev.\  D {\bf 58}, 061301 (1998)
  [astro-ph/9804177].

\bibitem{green}

  D.~R.~Green,
  Phys.\ Rev.\  D {\bf 76}, 103504 (2007)
  [arXiv:0707.3832].

\bibitem{preheat}

  J.~Braden, L.~Kofman and N.~Barnaby,
  JCAP {\bf 1007}, 016 (2010)
  [arXiv:1005.2196].

\bibitem{lorenzo2}

  M.~M.~Anber and L.~Sorbo,
  JCAP {\bf 0610}, 018 (2006)
  [astro-ph/0606534].

\bibitem{res}

  X.~Chen, R.~Easther and E.~A.~Lim,
  JCAP {\bf 0804}, 010 (2008)
  [arXiv:0801.3295].

\bibitem{Hannestad:2009yx}

  S.~Hannestad, T.~Haugbolle, P.~R.~Jarnhus and M.~S.~Sloth,
  JCAP {\bf 1006}, 001 (2010)
  [arXiv:0912.3527].










\bibitem{roulette}

  J.~R.~Bond, L.~Kofman, S.~Prokushkin and P.~M.~Vaudrevange,
  Phys.\ Rev.\  D {\bf 75}, 123511 (2007)
  [hep-th/0612197].

\bibitem{racetrack}

  J.~J.~Blanco-Pillado {\it et al.},
  JHEP {\bf 0411}, 063 (2004)
  [hep-th/0406230].

\end{thebibliography}
\end{document}